\documentclass[aps,prb,twocolumn,superscriptaddress,longbibliography,floatfix]{revtex4-2}

\usepackage{amsmath,amssymb,amsfonts,mathtools}
\usepackage{bm}
\usepackage{graphicx}
\usepackage[breaklinks=true,colorlinks=true,allcolors=blue]{hyperref}
\usepackage{microtype}
\allowdisplaybreaks[1]
\newcommand{\zetaE}{\zeta}
\newcommand{\Om}{\Omega}
\newcommand{\tausm}{\tau_{\rm sm}}
\newcommand{\taush}{\tau_{\rm sh}}
\newcommand{\tautr}{\tau_{\rm tr}}
\newcommand{\tauin}{\tau_{\rm in}}
\newcommand{\taupi}{\tau(\pi)}
\newcommand{\tauzero}{\tau(0)}

\begin{document}

\title{Amplitudes of Hall field-induced resistance oscillations with a two-harmonic density of states}

\author{Miguel Tierz}\email{tierz@simis.cn}\affiliation{Shanghai Institute for Mathematics and Interdisciplinary Sciences (SIMIS), Shanghai 200438, China}

\begin{abstract}
We derive explicit strong-field asymptotics for the normalized differential resistance in Hall field-induced resistance oscillations (HIRO) within the Vavilov--Aleiner--Glazman kinetic framework. For a single-harmonic density of states, the leading oscillation amplitude is set by the full backscattering rate $1/\taupi$. Extending the theory to a two-harmonic density of states, we show that the off-diagonal mixed kernel $\gamma_{12}$ admits an exact single-integral representation, from which the strong-field asymptotics follow directly. The resulting odd harmonics, notably $m=1$ and $m=3$, have coefficients determined by combinations of $1/\tauzero$ and $1/\taupi$, while the leading $m=2$ amplitude remains unchanged. On exact-kernel mock data generated and fit within the same model, with $\tautr$ and $\tauin$ held fixed, the resulting extraction protocol recovers $\tau_q$, $\taupi$, and---when the $m=1,3$ harmonics are resolved---$\tauzero$ to sub-percent accuracy, with $\tauzero$ providing a consistency check on the disorder description.
\end{abstract}

\maketitle

\section{Introduction}\label{sec:intro}
\enlargethispage{3\baselineskip}

\begin{figure*}[tb]
\centering
\includegraphics[width=0.85\textwidth]{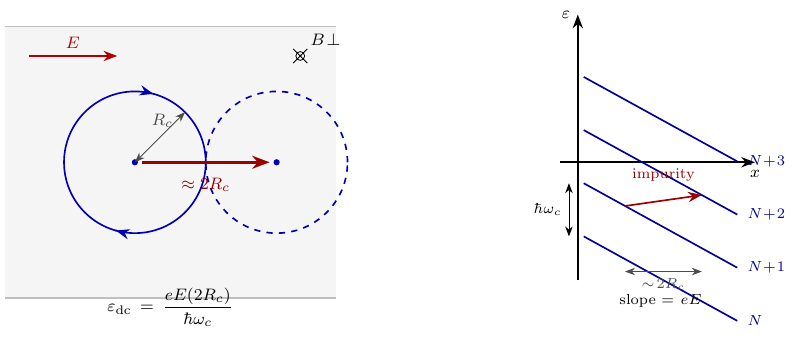}
\caption{HIRO geometry. Left: a cyclotron orbit of radius $R_c$ in a perpendicular field $B$ and a Hall field $E$ undergoes impurity-assisted guiding-center hops of order $2R_c$.  Right: the Hall field tilts the Landau levels; scattering between orbits separated by $\sim\!2R_c$ connects adjacent levels when $\varepsilon_{\mathrm{dc}}=eE(2R_c)/(\hbar\omega_c)$ is near an integer.}
\label{fig:hiro_schematic}
\end{figure*}

Hall field-induced resistance oscillations (HIRO) occur in high-mobility two-dimensional electron gases (2DEGs) when a dc Hall field $E$, together with impurity scattering, drives guiding-center displacements of order $2R_c$ between neighboring cyclotron orbits~\cite{YangPRL2002,BykovPRB2005,ZhangPRB2007}. The magnetoresistance then oscillates with the dimensionless parameter $\varepsilon_{\mathrm{dc}}=eE(2R_c)/(\hbar\omega_c)$, where $\omega_c=eB/m^*$ is the cyclotron frequency, and has extrema near integer values (Fig.~\ref{fig:hiro_schematic}). The closely related microwave-induced resistance oscillations (MIRO) share much of the same kinetic structure~\cite{Ryzhii1970,Mani2002,MIROexpts,ZudovReview}.

Vavilov, Aleiner, and Glazman~\cite{VAG2007} developed a microscopic kinetic theory of the dc nonlinear response, building on earlier work on the displacement mechanism (field-induced changes in the scattering phase space)~\cite{DmitrievMirlinPolyakov2003,DurstPRL2003,DmitrievMirlinPolyakov2005} and the inelastic mechanism (redistribution of the oscillatory part of the electron distribution)~\cite{VAG2004,KhodasVavilov2008}. In their formulation, the response is expressed in terms of two field-dependent kernels---$\Gamma_2$ for the displacement contribution and $\Gamma_1$ for the inelastic contribution---constructed from disorder-weighted sums of Bessel functions (Sec.~\ref{sec:framework}). Their analysis captured the parametric structure of the oscillations and the relevant crossover regimes, but left the overall prefactors implicit. In particular, the strong-field envelope approximation used in Eqs.~(18)--(19) of Ref.~\onlinecite{VAG2007} replaces the exact smooth-disorder sum by a nonoscillatory function and therefore discards exponentially small but finite oscillatory contributions from smooth backscattering.

These omissions matter when HIRO data are used as a disorder diagnostic. The oscillation amplitude is governed by the backscattering rate $1/\taupi$ (the rate for impurity scattering through angle $\pi$), the exponential damping by the quantum lifetime $\tau_q$, the low-field background by the inelastic time $\tauin$, and the baseline by the transport time $\tautr$~\cite{ZudovReview,DmitrievMirlinPolyakov2009}. In regimes where Shubnikov--de Haas (SdH) oscillations~\cite{AndoRMP1982} are thermally smeared---for example, by elevated temperatures or dc Joule heating---HIRO persist because their resonance condition is geometric rather than thermal~\cite{ZudovReview}. Fully explicit prefactors therefore provide, within the model and supplemented by baseline and temperature-dependent measurements, a route to constraining four scattering times ($\tautr$, $\tau_q$, $\taupi$, $\tauin$) from dc data. When the mixed $m=1,3$ harmonics are resolved, the forward-scattering rate $1/\tauzero$ (scattering through angle zero) becomes accessible as an additional consistency check of the disorder model~\cite{DmitrievMirlinPolyakov2009}. In practice, exact numerical evaluation of the kernels is preferable whenever the Dingle factor $\lambda=\exp[-\pi/(\omega_c\tau_q)]$ is not small (Sec.~\ref{sec:extraction}).

Another motivation comes from systems in which the Landau-quantized density of states (DOS) retains a visible second harmonic. This occurs in high-mobility GaAs/AlGaAs at moderate fields~\cite{ZudovPRB2017} and in MgZnO/ZnO heterostructures~\cite{ShiPRB2017_MgZnO}. Extending the single-harmonic theory of Ref.~\onlinecite{VAG2007} to a two-harmonic DOS introduces a mixed kernel $\gamma_{12}$ with unequal Bessel arguments, for which no closed-form product formula is available. We show that this kernel admits an exact single-integral representation (Appendix~\ref{app:toeplitz}), from which the strong-field asymptotics follow directly.

The remainder of this paper is organized as follows. Section~\ref{sec:framework} sets up the kinetic framework of Ref.~\onlinecite{VAG2007}. Section~\ref{sec:one-harmonic} evaluates the single-harmonic sums exactly and derives the explicit strong-field asymptotics. Section~\ref{sec:two-harmonic} extends the theory to a two-harmonic DOS. Section~\ref{sec:extraction} translates the results into a dc-only extraction protocol and validates it on synthetic data. Section~\ref{sec:discussion} discusses the broader context and outlook. The integral representation for $\gamma_{12}$ is derived in Appendix~\ref{app:toeplitz}.

\section{Kinetic framework}\label{sec:framework}

We work within the formulation of Ref.~\onlinecite{VAG2007}. Throughout, we use the notation
\begin{equation}
\tau_q\equiv \tau_0,\qquad \zetaE=\pi\varepsilon_{\mathrm{dc}},\qquad \lambda=e^{-\pi/(\omega_c\tau_q)}.
\label{eq:notation}
\end{equation}
Here $\tau_0\equiv\tau_q$ is the zeroth angular harmonic of the scattering rate. Despite the similar notation, the forward-scattering rate $1/\tauzero\equiv\sum_n\tau_n^{-1}$, introduced in Sec.~\ref{sec:two-harmonic}, is a distinct quantity.

The disorder potential contains short-range (sharp) and long-range (smooth) components with scattering rates $\taush^{-1}$ and $\tausm^{-1}$; the angular-harmonic rates are~\cite{VAG2007,DmitrievMirlinPolyakov2009}
\begin{equation}
\frac{1}{\tau_n}=\frac{1}{\tausm}\frac{1}{1+\chi n^2}+\frac{\delta_{n,0}}{\taush},
\label{eq:taun}
\end{equation}
where $\chi\ll 1$ parametrizes the smoothness of the long-range component. The dc Hall field enters through the dimensionless parameter $\zetaE\propto eER_c/(\hbar\omega_c)$. Averaging the phase factor $e^{i\zetaE(\sin\varphi'-\sin\varphi)}$ against $\tau^{-1}(\varphi'-\varphi)$ projects onto Bessel functions (Appendix~\ref{app:toeplitz}) and yields the disorder kernel~\cite{VAG2007}
\begin{equation}
\gamma(\zetaE)=\sum_{n\in\mathbb{Z}}\frac{J_n(\zetaE)^2}{\tau_n}.
\label{eq:gammadef}
\end{equation}
The displacement and inelastic kernels are~\cite{VAG2007}
\begin{align}
\frac{\Gamma_2}{\tautr}&=-\gamma''(\zetaE),\label{eq:Gamma2def}\\
\frac{\Gamma_1}{\tautr}&=-\frac{[\gamma'(\zetaE)]^2}{\tauin^{-1}+\tau_0^{-1}-\gamma(\zetaE)},\label{eq:Gamma1def}
\end{align}
and the kernel combination governing the nonlinear response is $F(\zetaE)=2\Gamma_1+\Gamma_2$. Our goal is to evaluate Eqs.~\eqref{eq:gammadef}--\eqref{eq:Gamma1def} exactly and then isolate the prefactors most relevant for experiment.

\section{Single-harmonic DOS: exact evaluation and asymptotics}\label{sec:one-harmonic}

The smooth-disorder contribution to Eq.~\eqref{eq:gammadef} is the weighted sum $\sum_n J_n^2(\zetaE)/(1+\chi n^2)$. Decomposing $(1+\chi n^2)^{-1}$ into partial fractions and applying Newberger's identity~\cite{Newberger1982}---which evaluates sums of the form $\sum_n J_n^2(x)/(n-\mu)$ in closed form as products of Bessel functions of complex order---with $\mu=\pm i/\sqrt{\chi}$ gives
\begin{equation}
\sum_n \frac{J_n^2(\zetaE)}{1+\chi n^2}=\frac{\pi}{\sqrt{\chi}\sinh(\pi/\sqrt{\chi})}\,J_a(\zetaE)\,J_{-a}(\zetaE),
\label{eq:newberger-exact}
\end{equation}
with $a=i/\sqrt{\chi}$, where $J_a$ denotes the analytic continuation of the Bessel function of the first kind to purely imaginary order~\cite{Watson1944,DLMF}. Adding the sharp contribution then yields $\gamma(\zetaE)$ in closed form.

Applying the Debye (large-argument) asymptotic expansion~\cite{Watson1944,DLMF} to the Bessel functions of both real and imaginary order, valid when $\zetaE\gg 1$, we obtain
\begin{align}
\gamma'(\zetaE)&=\frac{2}{\pi}\frac{1}{\taupi}\frac{\cos 2\zetaE}{\zetaE}+O(\zetaE^{-2}),\label{eq:gammaprime}\\
\gamma''(\zetaE)&=-\frac{4}{\pi}\frac{1}{{\taupi}}\frac{\sin 2\zetaE}{\zetaE}+O(\zetaE^{-2}),\label{eq:gammadoubleprime}
\end{align}
where the full backscattering rate
\begin{equation}
\frac{1}{\taupi}\equiv\sum_n \frac{(-1)^n}{\tau_n}=\frac{1}{\taush}+\frac{1}{\tausm}\frac{\pi}{\sqrt{\chi}\sinh(\pi/\sqrt{\chi})}
\label{eq:taupi-def}
\end{equation}
incorporates both the sharp-disorder contribution and an exponentially small smooth-disorder piece. The explicit strong-field kernels are
\begin{align}
\frac{\Gamma_2}{\tautr}&=\frac{4}{\pi}\frac{1}{\taupi}\frac{\sin 2\zetaE}{\zetaE}+O(\zetaE^{-2}),\label{eq:Gamma2-exact}\\
\frac{\Gamma_1}{\tautr}&=-\frac{1}{\tauin^{-1}+\tau_0^{-1}}\bigg(\frac{2}{\pi}\frac{1}{\taupi}\bigg)^{\!2}\frac{\cos^2 2\zetaE}{\zetaE^{2}}+O(\zetaE^{-3}).\label{eq:Gamma1-exact}
\end{align}
These expressions make the prefactors in Eqs.~(16)--(19) of Ref.~\onlinecite{VAG2007} fully explicit. The $\sin 2\zetaE/\zetaE$ form of the displacement kernel was first derived in Ref.~\onlinecite{DmitrievMirlinPolyakov2005}.

\begin{figure*}[tb]
  \centering
  \includegraphics[width=0.85\textwidth]{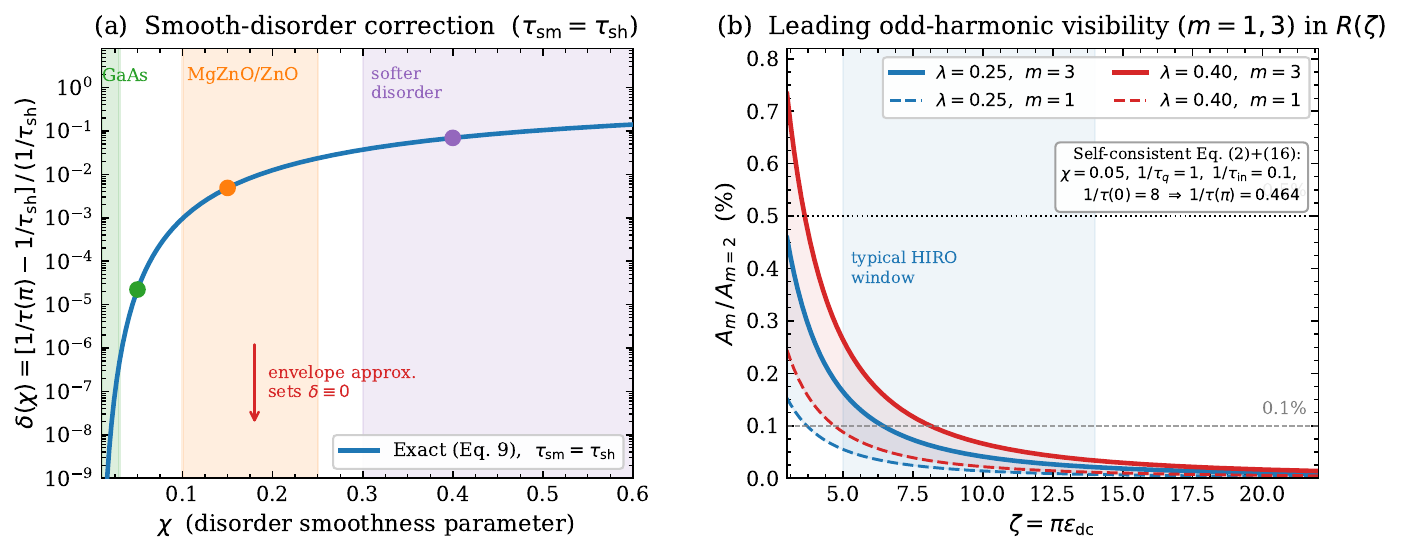}
  \caption{(a)~Smooth-disorder correction $\delta(\chi)\equiv[1/\taupi-1/\taush]/(1/\taush)$ to the leading HIRO oscillation amplitude, from Eq.~\eqref{eq:taupi-def} with $\tausm=\taush$ (the general dependence scales as $\taush/\tausm$). The envelope approximation of Ref.~\onlinecite{VAG2007} implicitly sets $\delta\equiv 0$. Shaded bands indicate representative disorder-smoothness ranges for GaAs, MgZnO/ZnO, and softer-disorder heterostructures (schematic). (b)~Leading odd-harmonic visibility $A_m/A_{m=2}$ in the measured response $\mathcal R(\zetaE)=d[\zetaE F]/d\zetaE$, from the asymptotic estimate of $\delta\Gamma_1^{(\mathrm{mix})}$, Eq.~\eqref{eq:Gamma1-mix}. Parameters are self-consistent with Eqs.~\eqref{eq:taun} and~\eqref{eq:tauzero-def} at $\chi=0.05$: fixing $1/\tau_q=1$ and $1/\tauzero=8$ (all rates in units of $1/\tautr$) yields $1/\taupi=0.464$. The asymptotic form becomes quantitatively reliable once $\chi\zetaE\gg 1$, i.e.\ $\zetaE\gg 1/\chi\approx 20$ at $\chi=0.05$. The shaded HIRO window $5\lesssim\zetaE\lesssim 14$ therefore sits in the crossover regime, where the asymptotic serves as an interpretive guide and the exact numerical kernel [Fig.~\ref{fig:gamma12_validation}] should be used for quantitative fitting. The factor-of-3 ratio $A_3/A_1=3$ is fixed by the derivative structure of Eq.~\eqref{eq:Gamma1-mix} and does not depend on the disorder parameters.}
  \label{fig:envelope}
\end{figure*}

The envelope approximation of Ref.~\onlinecite{VAG2007} replaces the exact smooth sum by $(1+\chi\zetaE^2)^{-1/2}$, retains only $\taush^{-1}$ in the oscillatory prefactor, and discards the second term in Eq.~\eqref{eq:taupi-def}. For typical GaAs parameters ($\chi\sim 10^{-2}$), this correction is exponentially suppressed [Fig.~\ref{fig:envelope}(a)], but it becomes non-negligible for softer disorder or for materials with larger $\chi$. In either case, Eqs.~\eqref{eq:Gamma2-exact}--\eqref{eq:Gamma1-exact} include it automatically. As a consistency check, expanding Eq.~\eqref{eq:newberger-exact} at $\zetaE\ll 1$ reproduces the weak-field scaling of Ref.~\onlinecite{VAG2007}.

\section{Two-harmonic DOS}\label{sec:two-harmonic}

We now retain the next harmonic of the Landau-quantized DOS in the standard Dingle hierarchy~\cite{AndoRMP1982}, writing $\nu(\varepsilon)=\nu_0[1-2\lambda\cos(\Om\varepsilon)+2\lambda^2\cos(2\Om\varepsilon)]$, where $\Om=2\pi/(\hbar\omega_c)$ is the inverse level spacing and $\lambda=\exp[-\pi/(\omega_c\tau_q)]$ is the single-harmonic Dingle factor [cf.\ Eq.~\eqref{eq:notation}]. The harmonic amplitudes $a_1=-2\lambda$ and $a_2=2\lambda^2$ are not independent but are fixed by Dingle damping; allowing arbitrary $a_2$ is straightforward but unnecessary for the Lorentzian level shape assumed here. The oscillatory isotropic distribution therefore takes the form $\delta f_{\rm iso}(\varepsilon)\propto a_1 I_1\sin(\Om\varepsilon)+a_2 I_2\sin(2\Om\varepsilon)$. Projecting the kinetic equation~\cite{VAG2007} onto $\{\sin\Om\varepsilon,\,\sin 2\Om\varepsilon\}$ yields a coupled $2\times 2$ system for the reduced amplitudes $(I_1,I_2)$, in which the DOS harmonic factors $a_k$ have been divided out so that the kernels $\gamma_{kj}$ carry no $\lambda$ dependence:
\begin{equation}
\begin{pmatrix}\Delta_{11} & -\gamma_{12}\\ -\gamma_{12} & \Delta_{22}\end{pmatrix}
\begin{pmatrix}I_1\\ I_2\end{pmatrix}
=-2eER_c\begin{pmatrix}\gamma'_{11}\\ \gamma'_{22}\end{pmatrix},
\label{eq:system-main}
\end{equation}
with $\Delta_{kk}=\tauin^{-1}+\tau_0^{-1}-\gamma_{kk}$ and kernels
\begin{equation}
\gamma_{kj}(\zetaE)=\sum_n \frac{J_n(k\zetaE)\,J_n(j\zetaE)}{\tau_n},\quad k,j\in\{1,2\}.
\label{eq:gammakj}
\end{equation}
The diagonal kernels $\gamma_{11}$ and $\gamma_{22}$ have equal Bessel arguments and are evaluated in closed form as in Sec.~\ref{sec:one-harmonic}. The off-diagonal kernel $\gamma_{12}$ has unequal arguments $(\zetaE,2\zetaE)$, for which no analogous closed form is available; its exact evaluation is given in Appendix~\ref{app:toeplitz} in the form of a single-integral representation:
\begin{align}
\gamma_{12}^{\rm sm}(x)&=\frac{1}{2\pi\,\tausm}\notag\\
&\quad\times\int_{-\pi}^{\pi}\!
S_\chi(\theta)\,
J_0\!\Big(x\sqrt{1+8\sin^2\!\tfrac{\theta}{2}}\Big)d\theta,
\label{eq:gamma12-main}
\end{align}
where $S_\chi(\theta)=(\pi/\sqrt{\chi})\cosh[(\pi-|\theta|)/\sqrt{\chi}]/\sinh(\pi/\sqrt{\chi})$ is the Fourier kernel of the weight $(1+\chi n^2)^{-1}$.

Stationary-phase evaluation of Eq.~\eqref{eq:gamma12-main} at $\theta=0$ and $\theta=\pi$ gives the strong-field asymptotics (Appendix~\ref{app:toeplitz}):
\begin{equation}
\gamma_{12}(\zetaE)=\frac{\sqrt{2}}{2\pi\,\zetaE}\bigg[\frac{\sin 3\zetaE}{\taupi}+\frac{\cos\zetaE}{\tauzero}\bigg]+O(\zetaE^{-2}),
\label{eq:gamma12-asympt}
\end{equation}
where $1/\tauzero\equiv\sum_n \tau_n^{-1}$ is the forward-scattering rate (not to be confused with the zeroth angular harmonic $1/\tau_0$), given in closed form by
\begin{equation}
\frac{1}{\tauzero}=\frac{1}{\taush}+\frac{1}{\tausm}\frac{\pi}{\sqrt{\chi}}\coth\!\left(\frac{\pi}{\sqrt{\chi}}\right).
\label{eq:tauzero-def}
\end{equation}

\begin{figure}[tb]
  \centering
  \includegraphics[width=\columnwidth]{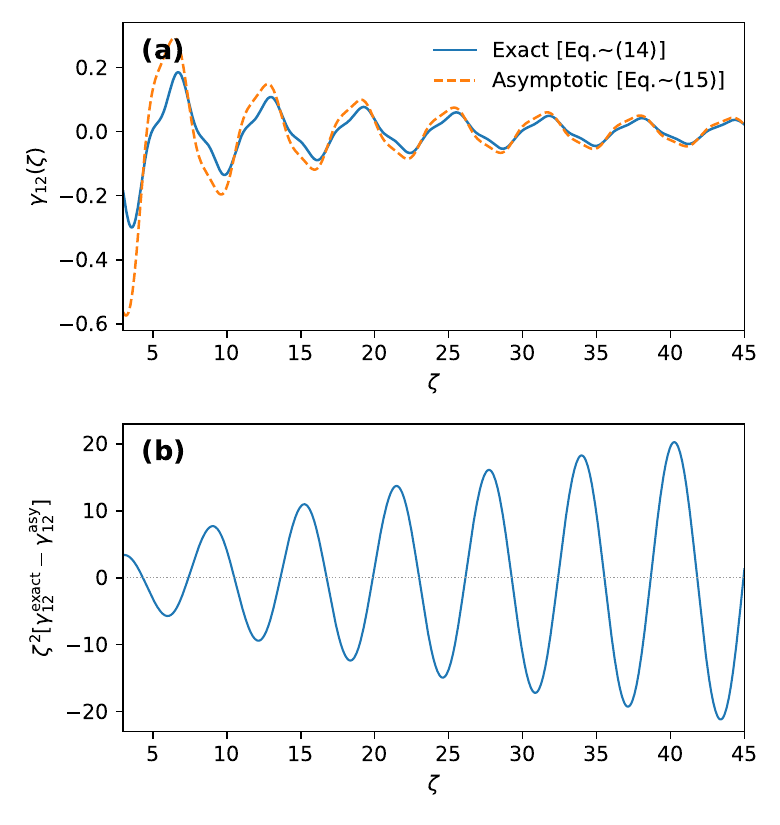}
  \caption{Validation of the mixed kernel $\gamma_{12}$.
  (a)~Exact evaluation of $\gamma_{12}$ (smooth integral plus sharp part $J_0(\zeta)J_0(2\zeta)/\taush$) and the strong-field asymptotic form Eq.~\eqref{eq:gamma12-asympt}.
  (b)~Scaled remainder $\zeta^2[\gamma_{12}^{\rm exact}(\zeta)-\gamma_{12}^{\rm asy}(\zeta)]$; the slow growth reflects the non-uniform $\chi\to 0$ behavior discussed in the text.
  Parameters are the same as in Fig.~\ref{fig:envelope}(b).}
  \label{fig:gamma12_validation}
\end{figure}

The two components have distinct frequency content: the $m=3$ term is weighted by $1/\taupi$, whereas the $m=1$ term is weighted by $1/\tauzero$. This separation in frequency space is the key diagnostic signature of the two-harmonic extension, and it leads directly to the odd-harmonic visibility shown in Fig.~\ref{fig:envelope}(b). Figure~\ref{fig:gamma12_validation} compares the exact evaluation of Eq.~\eqref{eq:gamma12-main} with the asymptotic form in Eq.~\eqref{eq:gamma12-asympt} and shows that the stationary-phase approximation reproduces the qualitative oscillatory structure of the mixed kernel in the field range relevant for $m=1,3$ harmonic analysis. The smooth-disorder part of Eq.~\eqref{eq:gamma12-asympt} is not uniform as $\chi\to 0$: the $1/\zetaE$ form becomes quantitatively reliable only once $\chi\zetaE\gg 1$ (equivalently $\zetaE\gg 1/\chi$), because the stationary-phase condition requires the Laplace-integral phase $\zetaE\chi t^2$ to oscillate appreciably over the range $t\sim 1$. At more moderate fields---including the typical HIRO window $5\lesssim\zetaE\lesssim 14$ at $\chi\sim 0.05$---the exact numerical kernel should be used for quantitative work, while the asymptotic form remains useful as an interpretive guide.

Inserting the asymptotics of all kernels into Eq.~\eqref{eq:system-main} and solving for the observables, we obtain the displacement kernel
\begin{align}
\frac{\Gamma_2}{\tautr}
&=\frac{4}{\pi}\frac{1}{\taupi}\frac{\sin 2\zetaE}{\zetaE}
+\frac{2}{\pi}\frac{1}{\taupi}\frac{\cos 2\zetaE}{\zetaE^{2}}
+\cdots,\label{eq:Gamma2-extended}
\end{align}
where the first term is the leading $m=2$ oscillation and the second is its subleading correction, both from the single-harmonic asymptotic of $-\gamma''(\zetaE)$. (A genuine $k=2$ DOS contribution from $\gamma_{22}(\zetaE)=\gamma(2\zetaE)$ oscillates at $m=4$, not $m=2$.) The ellipsis denotes additional $O(\zetaE^{-2})$ terms from the Debye correction to $J_n^2(\zetaE)$.

The off-diagonal kernel $\gamma_{12}$ enters the physical problem through the $2\times 2$ inelastic system~\eqref{eq:system-main}. Setting $\eta\equiv a_2/a_1=-\lambda$ for the ratio of the second and first DOS harmonic amplitudes, and noting that $|\eta|=\lambda<1$ for any partially resolved Landau spectrum, we solve Eq.~\eqref{eq:system-main} perturbatively in $\eta$: the feedback from $I_2$ into the first-harmonic amplitude $I_1$ is mediated by the off-diagonal element $\gamma_{12}$, so the mixed correction to $I_1$ enters at linear order in $\eta$. The resulting correction to the first-harmonic inelastic amplitude is
\begin{equation}
\frac{\delta\Gamma_1^{(\mathrm{mix})}}{\tautr}
=-\eta\,\frac{\gamma'_{11}(\zetaE)\,\gamma'_{22}(\zetaE)\,\gamma_{12}(\zetaE)}{A^2}
+O(\eta\,\zetaE^{-4}),
\label{eq:Gamma1-mix}
\end{equation}
with $A\equiv\tauin^{-1}+\tau_q^{-1}$. Substituting the large-$\zetaE$ forms of $\gamma'_{11}$, $\gamma'_{22}$, and $\gamma_{12}$ [Eqs.~\eqref{eq:gammaprime} and~\eqref{eq:gamma12-asympt}], the $m=1,3$ content is
\begin{align}
\frac{\delta\Gamma_{1,\,m=1,3}^{(\mathrm{mix})}}{\tautr}
&=-\frac{\eta\sqrt{2}}{2\pi^3 A^2\taupi^2}\,
\frac{1}{\zetaE^3}\bigg[
\frac{\sin\zetaE-\sin 3\zetaE}{\taupi}
\notag\\
&\quad+\frac{\cos\zetaE+\cos 3\zetaE}{\tauzero}\bigg]+\cdots,
\label{eq:Gamma1-mix-harmonics}
\end{align}
where the omitted terms at the same formal order carry $m=5,7,9$. Each of the $m=1$ and $m=3$ harmonics involves both $1/\taupi$ and $1/\tauzero$ with different phases; since $\taupi$ is already fixed by the leading $m=2$ amplitude, the residual $m=1,3$ content constrains $\tauzero$. Note that $\delta\Gamma_1^{(\mathrm{mix})}$ is $O(\lambda\,\zetaE^{-3})$ in $F$, but produces $O(\lambda\,\zetaE^{-2})$ content in the measured $\mathcal R(\zetaE)=d[\zetaE F]/d\zetaE$, so the mixed harmonics are experimentally accessible at moderate $\zetaE$.

Section~\ref{sec:extraction} provides an end-to-end numerical validation using exact-kernel mock data (Fig.~\ref{fig:recovery}).

For completeness, the one-harmonic inelastic kernel is
\begin{align}
\frac{\Gamma_1}{\tautr}
&=-\frac{1}{\tauin^{-1}+\tau_0^{-1}}\bigg(\frac{2}{\pi}\frac{1}{\taupi}\bigg)^{\!2}\frac{\cos^{2}2\zetaE}{\zetaE^{2}}
\notag\\
&\quad+\frac{1}{\tauin^{-1}+\tau_0^{-1}}\bigg(\frac{2}{\pi}\frac{1}{\taupi}\bigg)^{\!2}\frac{\sin 4\zetaE}{2\zetaE^{3}}+O(\zetaE^{-3}).
\label{eq:Gamma1-extended}
\end{align}
The leading oscillation $\propto\sin 2\zetaE/\zetaE$ in Eq.~\eqref{eq:Gamma2-extended} is unchanged: the second DOS harmonic does not renormalize the dominant amplitude $4/(\pi\taupi)$. In the measured signal $\Delta r/\rho_D=2\lambda^2\mathcal R(\zetaE)$ [Eq.~\eqref{eq:reduced-signal}], the Dingle factor $2\lambda^2$ multiplies the one-harmonic terms; the mixed $1\leftrightarrow 2$ pieces carry an additional factor of $\lambda$ relative to those, and the pure $2\leftrightarrow 2$ pieces an additional $\lambda^2$, so that the mixed harmonics become visible only when $\lambda$ is not small (partially resolved Landau levels)~\cite{ZudovPRB2017,ShiPRB2017_MgZnO}; the resulting $m=1,3$ visibility relative to the leading $m=2$ amplitude is shown in Fig.~\ref{fig:envelope}(b).

At low fields, expanding $J_n(k\zetaE)$ in powers of $\zetaE$ gives $\gamma_{11}=\tau_0^{-1}-(\zetaE^2/2)(\tau_0^{-1}-\tau_1^{-1})+O(\zetaE^4)$, $\gamma_{22}=\tau_0^{-1}-2\zetaE^2(\tau_0^{-1}-\tau_1^{-1})+O(\zetaE^4)$, and $\gamma_{12}=\tau_0^{-1}-(\zetaE^2/4)(5\tau_0^{-1}-4\tau_1^{-1})+O(\zetaE^4)$. Note that all three kernels share the same leading value $\tau_0^{-1}$; the off-diagonal coupling $\gamma_{12}$ is not small in $\zetaE$. The physical suppression of the two-harmonic correction arises from the Dingle factors: in the full signal $\Delta r/\rho_D=2\lambda^2\mathcal R$, the one-harmonic terms are $O(\lambda^2)$; the mixed $1\leftrightarrow 2$ correction [Eq.~\eqref{eq:Gamma1-mix}], proportional to $\eta=-\lambda$, enters at $O(\lambda^3)$; and the pure $2\leftrightarrow 2$ pieces at $O(\lambda^4)$. The leading weak-field behavior is therefore $\Gamma_2/\tautr=(\tau_0^{-1}-\tau_1^{-1})+O(\zetaE^2)$ and $\Gamma_1/\tautr=-\tauin(\tau_0^{-1}-\tau_1^{-1})^2\zetaE^2+O(\zetaE^4)$, with background shifts from the second harmonic entering at $O(\lambda^3)$.

\section{Extraction protocol}\label{sec:extraction}

The results of Secs.~\ref{sec:one-harmonic} and~\ref{sec:two-harmonic} provide a fitting framework for constraining $\tautr$, $\tau_q$, $\taupi$, and $\tauin$ from dc HIRO data, with the forward-scattering time $\tauzero$ additionally accessible when the mixed $m=1,3$ harmonics are resolved. Table~\ref{tab:hiro_mapping} summarizes the correspondence between observable features and microscopic scattering times.

To connect the theory directly to experiment~\cite{VAG2007,ZudovReview}, we write the nonlinear differential response in the reduced form
\begin{align}
\frac{\Delta r}{\rho_D}&=2\lambda^2\,\mathcal R(\zetaE),\qquad
\mathcal R(\zetaE)=\frac{d}{d\zetaE}\!\left[\zetaE\, F(\zetaE)\right],\notag\\
F(\zetaE)&=2\Gamma_1(\zetaE)+\Gamma_2(\zetaE),
\label{eq:reduced-signal}
\end{align}
with $\lambda=\exp[-\pi/(\omega_c\tau_q)]$ and $\zetaE=\pi\varepsilon_{\mathrm{dc}}$. The large-$\zetaE$ fitting form follows from Eqs.~\eqref{eq:Gamma2-extended}--\eqref{eq:Gamma1-extended} and~\eqref{eq:Gamma1-mix}:
\begin{align}
F(\zetaE)&=
\frac{4\tautr}{\pi\taupi}\,\frac{\sin 2\zetaE}{\zetaE}
+\frac{2\tautr}{\pi\taupi}\,\frac{\cos 2\zetaE}{\zetaE^2}
-\frac{8\tautr}{\pi^2 A\taupi^2}\,
\frac{\cos^2 2\zetaE}{\zetaE^{2}}
\notag\\
&\quad+2\,\delta\Gamma_1^{(\mathrm{mix})}
+\delta F_{\rm rem}(\zetaE)
+O(\zetaE^{-3}),
\label{eq:F-asymptotic}
\end{align}
with $A=\tauin^{-1}+\tau_q^{-1}$. The first two terms are the leading and subleading one-harmonic displacement contributions; the third is the inelastic envelope correction; and the fourth is the mixed inelastic correction from Eq.~\eqref{eq:Gamma1-mix}, which is $O(\lambda\,\zetaE^{-3})$ in $F$. The remainder $\delta F_{\rm rem}$ collects additional $O(\zetaE^{-2})$ contributions from the Debye correction to $J_n^2(\zetaE)$; it shares the $m=2$ frequency of the displayed terms and does not affect the $m=1,3$ harmonic content. Equation~\eqref{eq:F-asymptotic} is therefore exact for the $m=1,3$ structure but incomplete at $O(\zetaE^{-2})$ in the $m=2$ channel, so quantitative $m=2$ fitting should use the exact numerical kernels. Although the mixed piece is formally subleading in $F$, the derivative $\mathcal R(\zetaE)=d[\zetaE F]/d\zetaE$ raises it to $O(\lambda\,\zetaE^{-2})$ in the measured signal, so the $m=1,3$ content remains experimentally accessible. Explicitly, the leading $m=1,3$ part of the measured signal is
\begin{align}
\mathcal R_{m=1,3}^{(\mathrm{mix})}(\zetaE)
&=\frac{\lambda\sqrt{2}\,\tautr}{\pi^3 A^2\taupi^2}\,
\frac{1}{\zetaE^2}
\notag\\
&\quad\times\!\left[
\frac{\cos\zetaE-3\cos 3\zetaE}{\taupi}
-\frac{\sin\zetaE+3\sin 3\zetaE}{\tauzero}
\right]
\notag\\
&\quad+O(\lambda\,\zetaE^{-3}),
\label{eq:R-odd}
\end{align}
with $A=\tauin^{-1}+\tau_q^{-1}$, $\lambda=\exp[-\pi/(\omega_c\tau_q)]$, and the full measured odd-harmonic signal $(\Delta r/\rho_D)_{m=1,3}=2\lambda^2\,\mathcal R_{m=1,3}^{(\mathrm{mix})}$. This expression generates $m=1$ and $m=3$ Fourier content with distinct phase-resolved dependence on $1/\taupi$ and $1/\tauzero$; higher odd harmonics ($m=5,7,\ldots$) appear at the same formal order but are not displayed. The low-$\zetaE$ analysis must use the weak-field expansion quoted at the end of Sec.~\ref{sec:two-harmonic} rather than the large-$\zetaE$ asymptotic Eq.~\eqref{eq:F-asymptotic}.

\begin{figure*}[tb]
  \centering
  \includegraphics[width=0.85\textwidth]{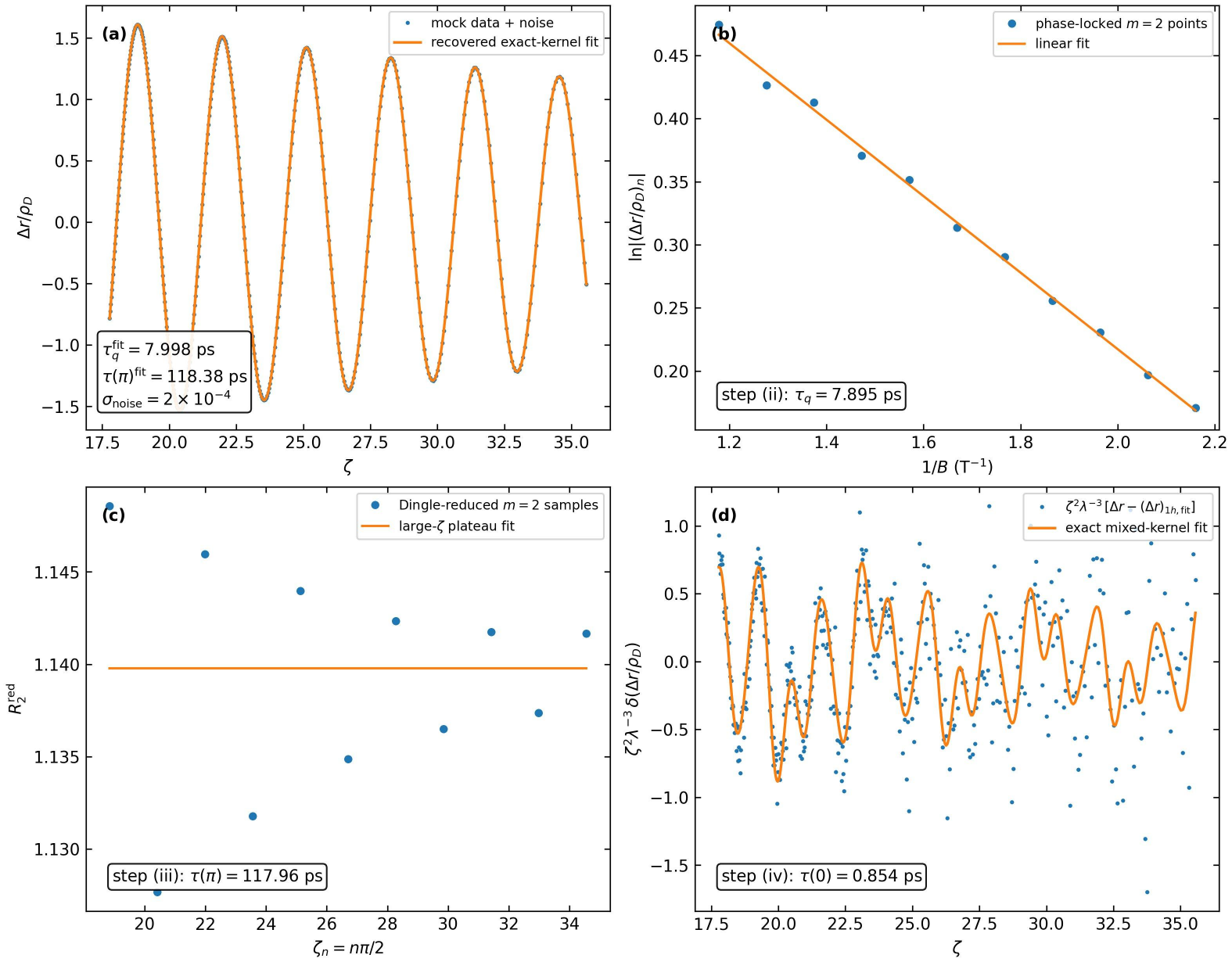}
  \caption{Synthetic recovery test. Mock data generated from exact numerical kernels [Eqs.~\eqref{eq:gammadef}--\eqref{eq:gamma12-main}] with Gaussian noise ($\sigma=2\times 10^{-4}$); true values $\tau_q=8.000$~ps, $\taupi=118.4$~ps, $\tauzero=0.857$~ps ($\tautr$ and $\tauin$ held fixed). (a)~Mock signal and recovered exact-kernel fit ($\tau_q^{\rm fit}=7.998$~ps, $\taupi^{\rm fit}=118.38$~ps). (b)~Dingle analysis [step~(ii)]: semilog plot of phase-locked $m=2$ extrema versus $1/B$ gives $\tau_q=7.895$~ps from the slope alone. (c)~Step~(iii): Dingle-reduced $m=2$ amplitude $\mathcal R_2^{\rm red}$ at successive extrema; the large-$\zetaE$ plateau determines $\taupi=117.96$~ps. (d)~Step~(iv): scaled odd residual $\zetaE^2\lambda^{-3}\,\delta(\Delta r/\rho_D)$ after subtracting the recovered one-harmonic baseline, fitted with the exact mixed-kernel template; the fit gives $\tauzero=0.854$~ps. The exact one-harmonic refit recovers $\tau_q$ and $\taupi$ to better than 0.1\%; the odd-harmonic fit recovers $\tauzero$ to 0.3\%.}
  \label{fig:recovery}
\end{figure*}

The extraction proceeds in five steps. Experimentally, one records current sweeps at fixed $B$, converts the current density to $\zetaE=\pi\varepsilon_{\mathrm{dc}}$, Fourier decomposes $\mathcal R(\zetaE)$ over a suitable $\zetaE$-window, and tracks the extracted $m=2$ and $m=1,3$ amplitudes as functions of $1/B$. In practice the window should span several full oscillation periods of the $m=2$ component while remaining within the asymptotic regime $\zetaE\gtrsim 5$; $\zetaE\in[5,20]$ is a practical working window for extracting the dominant $m=2$ amplitude. Quantitative fitting of the odd-harmonic content, which typically falls in the crossover regime $\chi\zetaE\lesssim 1$, should use the exact numerical kernels rather than the large-$\zetaE$ asymptotic forms. ``Phase-locked'' refers to evaluating the envelope at the discrete set of $\zetaE$ where the $m=2$ component has extrema [$\zetaE_n=n\pi/2$ for integer $n$; see Fig.~\ref{fig:recovery}(c)], which isolates the slowly varying envelope from the oscillatory phase.

(i) \emph{Geometry and baseline.}  Determine $\zetaE=\pi\varepsilon_{\mathrm{dc}}$ from the Hall-bar geometry and obtain $\tautr$ from the Drude baseline $\rho_D=m^*/(n_e e^2\tautr)$.

(ii) \emph{Quantum lifetime.}  Extract the $m=2$ Fourier component of $\mathcal R(\zetaE)$ and plot its amplitude on a semilog scale versus $1/B$; the slope determines $\tau_q$ [Fig.~\ref{fig:recovery}(b)].  When the two-harmonic correction is appreciable, the raw extrema of $\mathcal R$ are no longer a pure Dingle diagnostic, and one should instead use the Fourier-extracted $m=2$ component or a global fit to the full numerical kernels.

(iii) \emph{Backscattering rate.}  With $\tautr$ and $\tau_q$ fixed, fit the $m=2$ amplitude of $\mathcal R(\zetaE)$ at large $\zetaE$. Its value $8\tautr/(\pi\taupi)$ determines $\taupi$.

(iv) \emph{Forward-scattering rate, when $m=1,3$ harmonics are resolved.}  Subtract the one-harmonic baseline from $\mathcal R(\zetaE)$ and isolate the $m=1$ and $m=3$ Fourier components of the residual.  Equation~\eqref{eq:R-odd} gives these components, which scale as $O(\lambda\,\zetaE^{-2})$ in $\mathcal R$.  Each harmonic contains a different phase-resolved combination of $1/\taupi$ and $1/\tauzero$, modulated by the common prefactor $A^{-2}=(\tauin^{-1}+\tau_q^{-1})^{-2}$.  Because $\taupi$ is already determined in step~(iii), a joint fit of the $m=1,3$ amplitudes constrains $\tauzero$ together with $A$; if $\tauin$ is known independently, $\tauzero$ follows directly from the residual amplitude and phase. Within the mixed-disorder model of Eq.~\eqref{eq:taun}, $\tauzero$ is not an independent parameter but a derived combination of $\taush$, $\tausm$, and $\chi$, so the odd-harmonic measurement serves as a consistency check of the disorder description rather than as a fifth free parameter. Without resolved $m=1,3$ harmonics, $\tauzero$ is not independently accessible in the present framework.

(v) \emph{Inelastic time.}  With the elastic parameters fixed, determine $\tauin$ from the low-$\zetaE$ curvature using the weak-field expansion of Sec.~\ref{sec:two-harmonic} or, preferably, the full kernel evaluated numerically. The expected Fermi-liquid scaling $\tau_{\mathrm{in}}^{-1}\propto T^2$ remains a useful cross-check~\cite{ZudovReview}.

Steps~(ii) and~(iii) remain valid even when SdH oscillations are thermally suppressed, because the HIRO resonance condition is geometric and the envelope is Dingle damped rather than thermally damped. There is, however, a tension between the large-$\zetaE$ asymptotic regime, which requires $\zetaE\gg 1$, and the regime in which the second DOS harmonic is visible, which requires $\lambda$ not to be small. Since $\zetaE\propto j/B$ while $\lambda$ grows with $B$, the overlap window is not automatic: increasing the current density to enlarge $\zetaE$ also increases Joule heating, which directly affects $\tauin$ and potentially $\tau_q$. In practice, the analytic forms Eqs.~\eqref{eq:Gamma2-extended}--\eqref{eq:Gamma1-mix-harmonics} serve as an interpretive guide, whereas quantitative fitting over the full field range should use the exact numerical kernels from Eqs.~\eqref{eq:gammadef}--\eqref{eq:Gamma1def} and~\eqref{eq:gammakj}. Figure~\ref{fig:recovery} demonstrates this strategy on synthetic data generated and fit within the same model: the Dingle analysis supplies an initial estimate of $\tau_q$, the exact one-harmonic numerical refit sharpens $\tau_q$ and $\taupi$ to better than 0.1\%, and the odd-harmonic residual recovers $\tauzero$ to 0.3\%. In this test $\tautr$ and $\tauin$ are held fixed; full joint recovery of all five times from experimental data will require additional constraints, such as the Fermi-liquid $T^2$ scaling of $\tauin$.

\begin{table}[tb]
\caption{How HIRO observables constrain the microscopic scattering times.}
\begin{ruledtabular}
\begin{tabular}{p{0.38\columnwidth}p{0.54\columnwidth}}
Signal feature & Extraction\\\hline
Drude baseline $\rho_D$ & $\tautr$ from $\rho_D=m^*/(n_e e^2\tautr)$\\
HIRO positions vs.\ $\varepsilon_{\mathrm{dc}}$ & Geometry ($2R_c$); phase check\\
$m\!=\!2$ envelope vs.\ $1/B$ & $\tau_q$ from the Dingle slope\\
Leading $m\!=\!2$ amplitude of $\mathcal R(\zetaE)$ & $\taupi$ via $8\tautr/(\pi\taupi)$\\
Residual $m\!=\!1,3$ Fourier components & $\tauzero$ jointly with $A$; consistency check of disorder model\\
Low-$\zetaE$ curvature & $\tauin$; $T^2$ cross-check
\end{tabular}
\end{ruledtabular}
\label{tab:hiro_mapping}
\end{table}

\section{Discussion}\label{sec:discussion}

The results above place the predictions of Ref.~\onlinecite{VAG2007} on a fully quantitative footing and extend them, within the same kinetic framework, to a two-harmonic DOS. The leading nonlinear oscillation remains a direct probe of the backscattering time $\taupi$. Higher DOS harmonics do not renormalize that leading amplitude; instead, they generate $m=1,3$ harmonics whose coefficients are set by combinations of $1/\tauzero$ and $1/\taupi$, appearing at $O(\lambda\,\zetaE^{-2})$ in the measured $\mathcal R(\zetaE)$ [Eq.~\eqref{eq:Gamma1-mix-harmonics}]. Because $\taupi$ is already determined from the $m=2$ channel, this odd-harmonic content probes the disorder anisotropy ratio $\tauzero/\taupi$, inaccessible from $m=2$ alone~\cite{DmitrievMirlinPolyakov2009}, and provides a consistency check of the mixed-disorder model. The inelastic time $\tauin$ enters both the subleading one-harmonic envelope and the mixed odd-harmonic sector through the common prefactor $A^{-2}=(\tauin^{-1}+\tau_q^{-1})^{-2}$, so both channels constrain it. This two-harmonic regime is physically relevant when the Dingle factor $\lambda=\exp[-\pi/(\omega_c\tau_q)]$ is not small, so that the $\lambda^2$ component of the DOS contributes appreciably~\cite{ZudovReview}---a regime reached in the highest-mobility GaAs structures at moderate fields~\cite{HatkePRB2011,ZudovPRB2017,Wang2023} and in MgZnO/ZnO heterostructures~\cite{ShiPRB2017_MgZnO}; the same systems also host hydrodynamic charge transport~\cite{WangPRB2022,AlekseevSemina2025,AlekseevAlekseeva2025} and high-order cyclotron-resonance features probed optically~\cite{Savchenko2021,Savchenko2024}.

The analytic tools used here have a counterpart in magnetoplasmon physics: the Bessel-product sums $\sum_n J_{n+p}(x)J_{n+q}(x)/(n-\mu)$ that determine the HIRO kernels also enter the nonlocal conductivity governing collective (Bernstein-mode) absorption in a magnetized 2DEG~\cite{Kapralov2022PRB}; the viscous counterpart of these resonances was analyzed in Ref.~\onlinecite{Alekseev2018}. Newberger's identity~\cite{Newberger1982} solves both problems: it evaluates the HIRO kernel at imaginary $\mu=i/\sqrt\chi$ (disorder pole) and the Bernstein-mode residue at integer $\mu=n$ (cyclotron-harmonic pole). The asymptotics differ---turning-point behavior $J_n^2(n)\!\sim\! n^{-2/3}$ for Bernstein modes~\cite{Tierz2025BM} versus large-argument behavior $J_n^2(\zetaE)\!\sim\! 2/(\pi\zetaE)$ for HIRO---but the integral representation (Appendix~\ref{app:toeplitz}) used here for $\gamma_{12}$ extends the shared framework to unequal-argument sums in both settings. An analogous application of the Bessel summation formulas to optomechanical attractor diagrams was given in Ref.~\onlinecite{RussoTierz2025}.

Relevant experimental context includes the first HIRO observation~\cite{YangPRL2002}, systematic dc studies~\cite{BykovPRB2005,ZhangPRB2007}, combined ac+dc driving~\cite{HatkePRB2008}, high-field and high-order measurements~\cite{DaiPRL2010,HatkePRB2011,Wang2023}, direct separation of displacement and inelastic contributions~\cite{KhodasPRL2010}, density-tunable studies in GaAs~\cite{ZudovPRB2017} and MgZnO/ZnO~\cite{ShiPRB2017_MgZnO}, engineered disorder arrays~\cite{BartelsPRB2025}, and polarization-resolved MIRO~\cite{Savchenko2022}; see Ref.~\onlinecite{ZudovReview} for a broad review and Ref.~\onlinecite{AndoRMP1982} for background on the Landau-quantized DOS.

The present framework extends in a straightforward way to three or more DOS harmonics: each additional harmonic introduces new off-diagonal kernels $\gamma_{1k}$ with Bessel arguments $(\zetaE,k\zetaE)$, all exactly representable as single integrals via the same identity~\eqref{eq:toeplitz-unequal}. In practice, however, the $k$th harmonic enters with weight $\lambda^k$, so for typical Dingle factors the third and higher harmonics are suppressed by at least an order of magnitude relative to the second. The two-harmonic treatment should therefore capture the leading correction for currently achievable sample qualities; extending beyond it is possible but unlikely to be necessary until samples with $\lambda\gtrsim 0.5$ become routine.

The same displacement and inelastic mechanisms that produce HIRO also generate THz-induced magnetooscillations in graphene~\cite{Monch2020}, the direct analogue of MIRO in a Dirac system. Because graphene at THz frequencies can reach Dingle factors $\lambda\sim 0.3$--$0.5$, it is a plausible setting in which the two-harmonic corrections derived here could become experimentally testable, provided the parabolic-band assumption underlying the present model is extended to a Dirac spectrum.

To summarize: the one-harmonic kernel $\gamma_{11}$ and its derivatives are evaluated in closed form [Eq.~\eqref{eq:newberger-exact}]; the mixed kernel $\gamma_{12}$ is reduced to a single integral [Eq.~\eqref{eq:gamma12-main}]; the strong-field asymptotics [Eqs.~\eqref{eq:Gamma2-extended}--\eqref{eq:Gamma1-mix-harmonics}] expose the harmonic structure and scattering-rate dependence; quantitative fitting should use the exact numerical kernels from Eqs.~\eqref{eq:gammadef}--\eqref{eq:gammakj}; and the synthetic recovery test (Fig.~\ref{fig:recovery})---based on exact-kernel mock data generated and fit within the same model, with $\tautr$ and $\tauin$ held fixed---confirms that the protocol recovers $\tau_q$, $\taupi$, and $\tauzero$ at the sub-percent level.


\bibliographystyle{apsrev4-2}
\bibliography{hiro_bib}

\appendix
\section{\texorpdfstring{Integral representation for $\gamma_{12}$}{Integral representation for gamma\_12}}\label{app:toeplitz}

The Bessel function of the first kind, $J_n(x)$, is defined for integer order $n$ by the Jacobi--Anger expansion $e^{ix\sin\theta}=\sum_n J_n(x)\,e^{in\theta}$~\cite{Watson1944,DLMF}. The HIRO kernels are built from sums of products of these functions weighted by the angular-harmonic scattering rates $1/\tau_n$ [Eq.~\eqref{eq:taun}].

When both Bessel functions share the same argument, as in $\gamma_{11}$ and $\gamma_{22}$, the sum $\sum_n J_n^2(x)/(n+\mu)$ can be evaluated in closed form as a product of Bessel functions of complex order~\cite{Newberger1982}, subject to $\mathrm{Re}(p+q)>-1$ for the shifted-index generalization $\sum_n J_{n+p}(x)J_{n+q}(x)/(n+\mu)$. For the mixed kernel $\gamma_{12}$, however, the arguments are $x$ and $2x$, and no such product formula is available.

The key tool is the following integral representation, which we now derive. The starting point is the Neumann addition theorem for $J_0$:
\begin{equation}
J_0\!\Big(\!\sqrt{x^2\!+\!y^2\!-\!2xy\cos\theta}\Big)
=\sum_{n=-\infty}^{\infty}J_n(x)\,J_n(y)\,e^{in\theta}.
\label{eq:neumann-addition}
\end{equation}
To see this, write
\begin{align}
\sum_n J_n(x)\,J_n(y)\,e^{in\theta}
&=\frac{1}{2\pi}\!\int_{-\pi}^{\pi}\!
e^{ix\sin(\theta-\psi)+iy\sin\psi}\,d\psi,
\end{align}
where we used the Jacobi--Anger expansion twice and the orthogonality of $\{e^{in\psi}\}$. The exponent can be written as $R\sin(\psi+\phi)$ with $R=\sqrt{x^2+y^2-2xy\cos\theta}$ and a $\theta$-dependent phase $\phi$; the $\psi$-integral then yields $J_0(R)$ by its standard integral representation.

Now let $S_w(\theta)=\sum_n w(n)\,e^{in\theta}$ be the Fourier series of any weight. Multiplying Eq.~\eqref{eq:neumann-addition} by $S_w(\theta)/(2\pi)$ and integrating over $\theta\in[-\pi,\pi]$, Fourier orthogonality selects the $n$th coefficient on both sides:
\begin{align}
&\sum_{n=-\infty}^{\infty} w(n)\,J_n(x)\,J_n(y)\notag\\
&\quad=\frac{1}{2\pi}\int_{-\pi}^{\pi}\! S_w(\theta)\,
J_0\!\Big(\!\sqrt{x^2\!+\!y^2\!-\!2xy\cos\theta}\Big)d\theta.
\label{eq:toeplitz-unequal}
\end{align}
The representation is termed ``Toeplitz'' because the weight enters only through the angular difference $\theta=\varphi-\varphi'$, so that the resulting matrix has entries depending only on the difference of their indices.

With the smooth-disorder weight $w(n)=(1+\chi n^2)^{-1}$, the Fourier kernel is
\begin{equation}
S_\chi(\theta)=\frac{\pi}{\sqrt{\chi}}\,\frac{\cosh[(\pi-|\theta|)/\sqrt{\chi}]}{\sinh(\pi/\sqrt{\chi})}
\end{equation}
for $|\theta|\le\pi$. Setting $y=2x$ gives Eq.~\eqref{eq:gamma12-main}; the sharp part is $\gamma_{12}^{\rm sh}(x)=J_0(x)J_0(2x)/\taush$.

Equations~\eqref{eq:toeplitz-unequal} and~\eqref{eq:gamma12-main} place $\gamma_{12}$ on the same analytic footing as $\gamma_{11}$ and $\gamma_{22}$, which admit parallel Toeplitz forms with the same kernel $S_\chi(\theta)$. The large-$x$ asymptotics follow from stationary phase at $\theta=0$ and $\theta=\pi$: the $\theta=\pi$ saddle point produces the $\sin 3x$ component weighted by $1/\taupi$, and $\theta=0$ produces the $\cos x$ component weighted by $1/\tauzero$, yielding Eq.~\eqref{eq:gamma12-asympt}.

An equivalent single-integral representation follows from $(1+\chi n^2)^{-1}=\int_0^\infty e^{-t}\cos(\sqrt\chi\,n t)\,dt$:
\begin{equation}
\gamma_{12}^{\rm sm}(x)=\frac{1}{\tausm}
\int_0^\infty \!e^{-t}\,
J_0\!\Big(x\sqrt{1+8\sin^2\!\tfrac{\sqrt{\chi}\,t}{2}}\Big)dt,
\label{eq:gamma12-laplace}
\end{equation}
which is useful for numerical evaluation.

An alternative packaging of the two-harmonic problem uses the generalized two-variable Bessel function $J_n(x,y)$~\cite{Dattoli1990}---a distinct mathematical object from the standard single-argument $J_n(x)$ used throughout this paper---defined by $e^{i[x\sin\theta+y\sin 2\theta]}=\sum_n J_n(x,y)\,e^{in\theta}$. In that notation the kernel takes the formally identical form $\gamma(x,y)=\sum_n J_n(x,y)^2/\tau_n$. This is convenient for bookkeeping and for the low-field expansion, because the perturbative coefficients of $J_n(x,y)$ in powers of $y$~\cite{Dattoli1990} reduce each order to standard shifted-index sums evaluable via Ref.~\onlinecite{Newberger1982}.

For the full problem, however, the closed-form product formula does not extend to sums of generalized Bessel functions. Expanding $J_n(x,y)=\sum_m J_{n-2m}(x)J_m(y)$ and applying the identity term by term violates its convergence condition for $m\ge 1$, producing an error of the same order as the leading oscillation. The Toeplitz representation~\eqref{eq:toeplitz-unequal} avoids this difficulty entirely.

\end{document}